\documentclass[runningheads]{llncs}
\usepackage[T1]{fontenc}
\usepackage{graphicx}
\begin{document}
\title{Dispersion of personal spaces}
\author{Jaroslav Hor\'{a}\v{c}ek\inst{1}\orcidID{0000-0002-0672-4339} \and
Miroslav Rada\inst{2,3}\orcidID{0000-0002-1761-897X}}
\authorrunning{J. Hor\'{a}\v{c}ek and M. Rada}
\institute{Charles University, Faculty of Humanities, Department of Sociology, Prague, Czech Republic \\
	\email{jaroslav.horacek@fhs.cuni.cz} \and
	Prague University of Economics and Business, Faculty of Informatics and Statistics, Department of Econometrics, Prague, Czech Republic \and
	Prague University of Economics and Business, Faculty of Finance and Accounting, Department of Financial Accounting and Auditing, Prague, Czech Republic\\
	\email{miroslav.rada@vse.cz}}
\maketitle              % typeset the header of the contribution
\begin{abstract}
There are many entities that disseminate in the physical space -- information, gossip, mood, innovation etc. Personal spaces are also entities that disperse and interplay.    	
In this work we study the emergence of configurations formed by participants when choosing a place to sit in a rectangular auditorium.   Based on experimental questionnaire data we design several models and assess their  relevancy to a real time-lapse footage of lecture hall being filled up. The main focus is to compare the evolution of entropy of occupied seat configurations in time. 
Even though the process of choosing a seat is complex and could depend on various properties of  participants or environment, some of the developed models can capture at least basic essence of the real processes. After introducing the problem of seat selection and related results in close research areas, we introduce preliminary collected  data and build models of seat selection based on them. We compare the resulting models to the real observational data and discuss areas of future research directions.   	

\end{abstract}

\section{Introduction}

Most of us have encountered a situation when we entered a lecture room or auditorium for the first time and had to choose a seat. A few people had been already seated there and we expected an indefinite number of people to arrive after us.  

It is very interesting to observe formation of such patterns from the perspective of a teacher or speaker. For example, at the beginning of his book   \cite{schelling2006micromotives} Schelling contemplates  a phenomenon that occurred during his lecture -- several front rows of seats were left vacant whilst  the other rows in an auditorium were fully packed.

Despite its common occurrence, phenomena corresponding to patterns resulting from seating preferences are not frequently studied. There is only few studies dedicated to this topic.  In \cite{bergtold2019spatial} they study the influence of seating on study performance.  The study  \cite{harms2015take} shows that students are biased to choose seats on the left side of the classroom. In \cite{haghighi2012exploring} they review the impact of seating arrangement.

An important factor influencing the seat selection process is territoriality. 
 The study  \cite{kaya2007territoriality} identified seats that are considered highly territorial -- seats at the end of rows or in the corners of a classroom.  It was showed that highly territorial seats attract students with greater territorial needs.
 Various arrangements of an auditorium were examined -- auditorium consisting of rows of tables with chairs, U-shaped auditorium, rows of tablet-arm chairs or chairs clustered around isolated tables.  
  
In  another study \cite{guyot1980classroom} they consider  interplay of human territoriality and seating preference.
They discovered that after several lectures students tend to occupy the same seat (or at least location). More than 90\% of students confirmed that they tend to always occupy the same location. However, they were unsure about the reasons for such a behavior. The following reasons received the highest scores -- security, sense of control or identity. However, there are several other reasons for this behavior such as social order, social role, stimulation etc. 

Human territoriality might have several meanings unlike the territoriality of animals \cite{guyot1980classroom}.  One of the possible meanings could be privacy. It is useful to understand it as control of access to other people. 
In \cite{pedersen1994privacy} they use Privacy Preference Scale (PPS)  that is based on this definition. It distinguishes various types of privacy -- intimacy, not neighboring, seclusion, solitude, anonymity, reserve.
Particularly seclusion and solitude need some further explanation. Seclusion is understood as being far from others, whilst solitude means being alone with others nearby.  All variants of privacy might be related to seating preference. In this study, rows in a lecture hall were marked as front, middle and back. Then various privacy scores of participants seated in those regions were measured. Most of the privacy factors were more important for participants seated towards the back. Nevertheless, some of the privacy conditions could be reached independently of seating in the back or front, e.g., seclusion, anonymity or not neighboring.

There are other studies focused on seating outside the lecture hall. 
Choosing a seat in a crowded coffee-house is studied in \cite{staats2019seat}. 
In his thesis \cite{thomas2009social} Thomas focuses on various situations in public transport. One section is devoted to spaces between passengers. There are documented examples of maximization of distances among passengers. If necessary, people tend to sit next to people expressing similarity (e.g., similar gender) or passengers not expressing a defensive behavior (e.g., placing a body part or other item on adjacent seat). The average distance of two passengers was measured to be 1.63 seat units. 
Another work apply  agent-based models to study emergence of clusters in a bus \cite{alam2008studying}. 

Most of the mentioned studies are descriptive -- they try to statically capture the properties of seating arrangements or participant deliberation about seat selection.  The purpose of this work is to design dynamic models of seat selection that can shed new light on the formation process of collective seating patterns. In the next section we describe what factors do we consider when building such models. In the third and fourth section we describe the preliminary experimental data and the setup for comparing models respectively. The next section is devoted to designing the models and their comparison. In the final section we summarize the results and outline future research directions.

\section{Factors to consider}

Let us imagine that a person (or a group of people) arrives in an auditorium and seeks a suitable place to sit. We will call such a person \emph{participant}.  
 
There are several factors that could influence participant's seat preference. One of them could be visual or olfactory attractiveness of already seated participants. We were often asked about these assumptions during experimental phase of this work. They could play a significant role, however we assume that all participants are basically neutral in appearance.  We are interested in situations when there is no such disruptive element in the auditorium.  

There are several types of influence outside the environment. It makes a difference if we know the speaker or whether  we  expect to be asked questions during a lecture. That could have a significant influence on our seat selection. Hence we assume that there is no familiarity or interaction during a lecture. 

Another factor we encountered in our experiments is the length of a lecture. Some of the participants would choose to sit at the ends of seat rows. Either because of having an escape way if the lecture is not satisfactory or to have easy access to rest rooms.

Two other factors are mentioned in \cite{schelling2006micromotives}. First, people might have learned the seating habits from other activities or events. For example, they have had a bad experience with sitting in the front.
The second is a hypothesis that people just do not think about the seat selection too much. They just take any near vacant seat, that satisfies some basic social rules, e.g. not too many people need to stand up when getting to a seat. The book suggest that it could be due to the of interest. However, we suggest that the reason could be  the limited time of deliberation on seat selection.  We return to this idea when designing seat selection models. 

In the introductory section the two important factors were mentions -- territoriality and privacy. We will definitely make use of them in the following sections.  

Seating preferences could be based on sensory capabilities of a participant. For example, a participant with hearing or vision impairment would rather sit in the front.  For now, we do not include these conditions in our models.  

In \cite{alam2008studying} they consider the following parameters of participants -- age, gender, size, ethnicity. We do not take these parameters into account either since results on relation of these parameters and seating preferences are not apparent. In \cite{kaya2007territoriality} they observed a higher territoriality with women, however, it could be explained simply by the fact that observed woman participants carried more items and hence needed more personal space.   

Another characteristics more common for public transport is shifting seats as mentioned in \cite{thomas2009social}. 
If ones seat is not satisfactory it is reasonable to change it. In real situations we can observe people switching their seats with somebody else or moving to a different location. Such a behavior is not included in our models but it is a subject of  future work.   

\section{Data and its evaluation}
To obtain the first results we rely on preliminary empirical data. As the first source we used  publicly available time lapse videos of lecture halls being filled up. We were able to find several videos, however, one represented the process of seating selection in a medium sized lecture hall particularly well\footnote{At the time of writing the paper (14 October, 2023), the footage remains publicly accessible at the link https://www.youtube.com/watch?v=r76dclwZU9M}. 

In the footage,  participants arrive alone or in small groups, hence we also consider adding new participants in bunches. This footage serves as the first inspiration for further research, because it neatly illustrates the formation of seating patterns. 

We followed the video until we were still able to track the order of seating of the participants. 
At that moment the majority of seats still remained unoccupied.  
Nevertheless, such a moment, when the auditorium is still empty enough, poses the most interesting case. When the auditorium is nearly full, then the seating preference is overpowered by the need of a seat. 

\begin{figure}
	\centering
	\includegraphics[width=6cm]{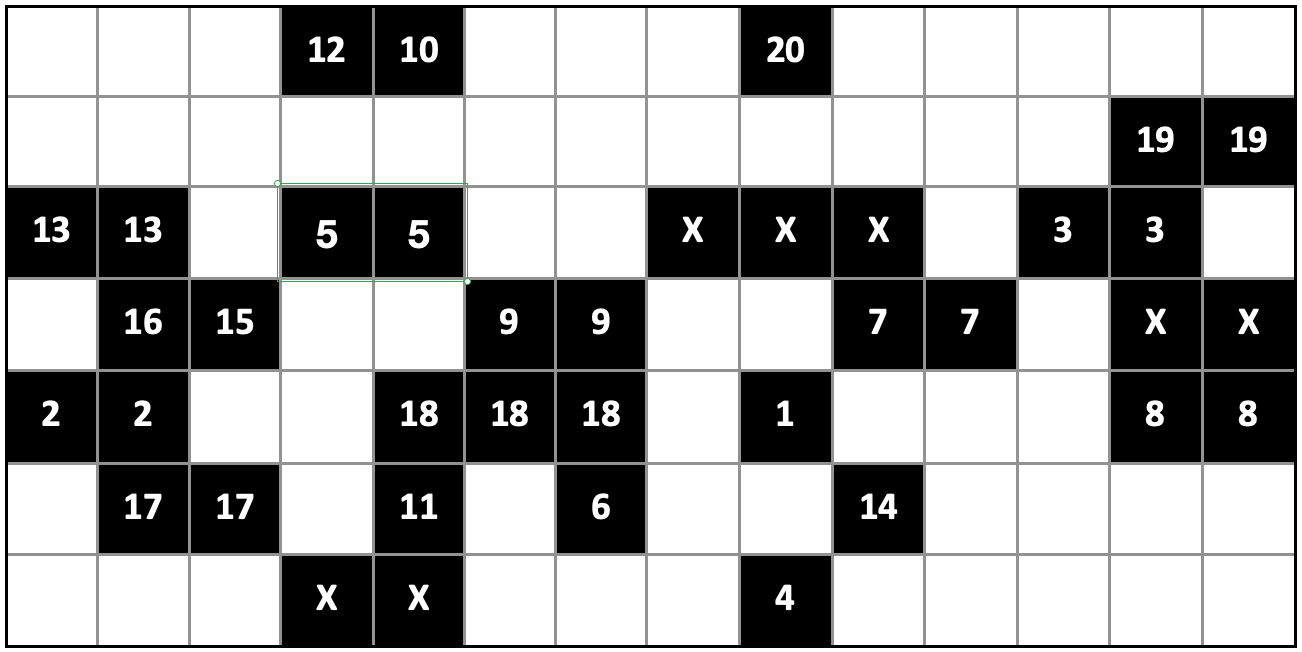}
	\caption{A tabular depiction of the real auditorium filled with participants. The black cells represent occupied seats. The white cells represent empty seats. The seats marked with X represent  participants seated before the start of recording. The number in the black cell depicts the order of arrival of the corresponding participant. } \label{fig:realseats}
\end{figure}

After extracting  the visible part of the auditorium, there remained 7 rows of seats. Each row contained 14 seats. Therefore, the auditorium can be represented by a rectangular grid of size $7 \times 14$, where each cell represents one seat. In Figure \ref{fig:realseats}  we show such a representation. The black cells represent occupied seats. The numbers in black cells denote the order of arrival of corresponding participants. If more cells share the same number it means that all these participants arrived at the same time as a group. 

To understand the process of seating pattern formation we took the same approach as in the paper \cite{staats2019seat}. We prepared and printed four schematic depictions of seating arrangements in  a $7 \times 14 $ auditorium with some already seated participants. It was represented in the tabular way mentioned above. The examples were based on the initial seating arrangements from the real footage. The preliminary data collection took two rounds. Several undergraduate university students were asked whether they want to participate in the following short thought experiment. All respondents participate in ICT study programs. The collection of data was voluntary and anonymous. We did not collect any personal information such as age, gender etc, since there is no proven influence of these factors on seating preference. 
   
In the first round, 12 students were asked about 4 seating situations depicted in Figure \ref{fig:realseats}. They were asked to mark their preferred seating position. If several positions seemed equivalent, they were asked to mark all of them. The first round served as a pretest of quality of the example seating configurations and we used it also to improve the overall experiment setup. 

\begin{figure}
	\centering
	\includegraphics[width=10cm]{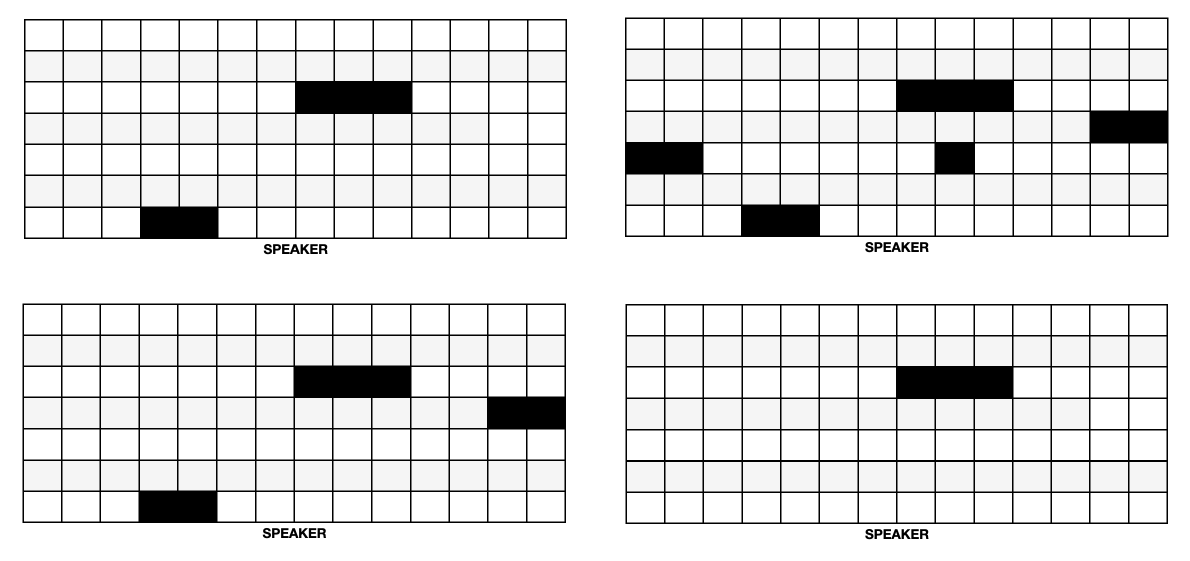}
	\caption{Four initial example configurations. The position of a speaker is marked for each hypothetical auditorium tableau. } \label{fig:testconf}
\end{figure}

In the second round there participated 27 students (not participating in the first round). The students were presented the same four hypothetical situations as in the first round, since they proved to be useful. This time, they we asked to select only one position of their preferred seat. They were instructed as follows: \emph{"Imagine you being a participant at a lecture. It is a one-time event. You participate voluntarily and are interested in the content of the lecture. You arrive from the back of the auditorium and need to pick one seat. Both sides of the auditorium are well accessible. You do not know anybody of the participants. The lecture is not long (maximum duration is 1 hour). There is no expectation of you being asked questions by a speaker during the lecture. Some people are already seated in the auditorium. They all seem neutral in appearance. You should definitely expect some more people to come. However, you do not know the amount of the newcomers. Which is your most preferred seat?"}   

Out of the 27 respondents, 9 marked exactly the same spot for all four configurations. There are several possible reasons for this outcome. First, they were bored with the task and hence filled the form without paying attention. Second, the seat choice depended on a preferred seat position in general and was hence independent of the seating arrangement (this was verbally confirmed by one of the participants). Anyway, in this preliminary testing phase we temporarily leave the 9 respondents out of our analysis since we do not know anything about their motivation fur such answers.  Interestingly, 8 of these 9 respondents chose a position at the edge of the auditorium.  

For our further analysis we work with 18 valid questionnaires. Since each consisted of 4 configurations, it gives us 72 cases of marked seating preferences in total.  

There are several interesting outcomes of the analysis. First, if there is enough space, respondents rarely choose a position directly neighboring (vertically, horizontally or diagonally) to someone. Such a  choice was made by only 3 participants in only 6 cases (about 8\% of all the cases).  

We first worked with the hypothesis that participants choose a seat that maximizes the distance to the nearest already seated participant. 
 Since the grid is discrete and rectangular, we used the Manhattan distance for such measurement. The set of all participants is denoted by $P$ and for a participant $p$ we denote $p_r, p_s$ his/her row and seat respectively. Then the distance of two participants $p, q$ is defined as:
{\small $$ d(p, q) = |p_r - q_r| + |p_s - q_s|.$$}

\begin{figure}
	\centering
	\includegraphics[width=6cm]{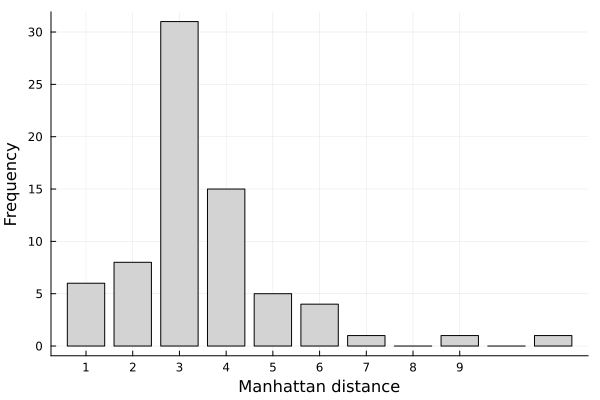}
	\caption{Frequency of distances to the nearest seated participant.} \label{fig:distances}
\end{figure}

Figure \ref{fig:distances} shows frequencies of distances to the nearest seated participant. It seems that participants tend to pick a seat close to already seated participants. However, some small gap is preferred (minimum distance equal to 3 appeared in 43\% of cases). 

Contrary to our hypothesis, participants do not seem to maximize their minimum distance to others. Only in 4 out of 72 cases could we see a seating choice that visually resembled maximization of distance to others. In the discarded set of participants a seat selection similar to maximization is more common (12 out of 36 cases). Nevertheless, maximization of distance to other participants is a phenomenon we return to in the next sections.  

Another way to measure  distance  is the distance from the center of the mass of already seated participants.
We assign numbers to the rows starting from the back (the largest number is of the  front row). Similarly, we number the seats from left to right.  Then, row $c_r$ and seat $c_s$ of the center of mass $c$ is defined as:
{\small $$c_r = \textrm{round} \left( \frac{\sum_{p \in P} p_r}{|P|} \right), \quad c_s = \textrm{round} \left( \frac{\sum_{p \in P} p_s}{|P|} \right).$$}
 We use rounding to the nearest integer with round half up tie-breaking rule. 
 
\begin{figure}
	\centering
	\includegraphics[width=6cm]{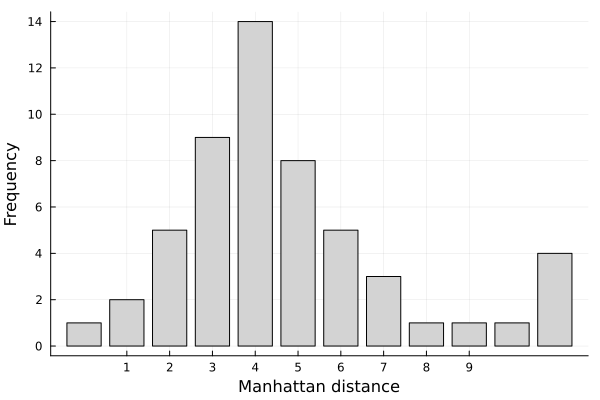}
	\caption{Frequency of distances to the center of mass. } 
	\label{fig:centerdists}
\end{figure}

 In Figure \ref{fig:centerdists} we show the frequency of distances to the center of the mass of already seated participants. For measurement we omitted the lower right configuration from Figure \ref{fig:testconf}  because there was only one seated groups. The largest distances were mostly produced by one respondent. It seems that a reasonable distance from the center of mass plays (maybe subconsciously) a role in the choice of a seat. We will utilize the results in these following sections. 

\section{Basic setup}
Using the results from the previous section, it is our goal to design models that capture the dynamics of seating formation. The data obtained from the real footage serves for comparison of developed models. 

For such a purpose the data from Figure \ref{fig:realseats} were slightly modified. The cells where participants sit next to each other and have different numbers are marked as one group of participants. From the footage it is clear that when a participant sat directly next to another one, they were acquainted and basically had intended to sit next to each other from the beginning. 

The general seating process is defined as follows. In one time step a group of participants arrives  and chooses a space where all group members can sit next to each other. In the real world participants could leave empty seats among them in the case when the auditorium is not fully packed. Since the participants do not know the number of participants arriving after them, we do not consider such a possibility and we always perceive a group as one unit that needs to be placed compactly.

Every simulation run shares the same initial configuration of already seated participants that is defined by "x" marks in Figure \ref{fig:realseats}. We use the same size of the auditorium ($7 \times 14$).    
That means that a group of 2, 2 and 3 participants respectively is already seated in the auditorium at the prescribed positions.  Next, we add groups of exactly the same sizes and in the same order  as in the real footage. We always assume that there is at least one spot where a group can fit. 

At every time step a configuration of already seated participants forms a certain pattern. To capture the essence of entropy  of such a pattern we devised the following simple measure. For every row it counts the number of transitions from an empty seat to a seated person and vice versa. For every row this count is squared and then these values are summed up. This is done to emphasize the rows of auditorium. This way an arrangement of  densely packed seated participants will obtain a small score whereas participants loosely spread across the auditorium with plenty of free space among them will get significantly higher score. We can represent an auditorium as a binary matrix $A$ where $A_{i, j}=1$ if the $j$-th seat in the row $i$ is occupied.  The matrix element $A_{i, j}$ is equal to 0 when the corresponding seat is empty.  For an auditorium of size  $m \times n$  the entropy measure is calculated as
{\small $$ e(A) = \sum_{r=1}^m \left(\sum_{s=2}^n | A_{r, s-1} - A_{r, s}|\right)^2.$$}

\textbf{Example:} \emph{Just one row with 14 occupied seats has entropy 0. One row where empty and occupied seats alternate has entropy 169. 
The entropy of the final seating arrangement from the real footage is 231.}

By measuring seating  entropy after arrival of each group we can capture the evolution of the entropy in time. 
All tested models are stochastic -- they use randomness to choose among equivalently good seating positions. Therefore, for each model  of seat selection the evolution of entropy in time is obtained by averaging  1,000 independent runs of the model.  

\section{Models of seat selection and their comparison}
Using the analysis of the empirical data we design several models of seating behavior and compare them with the real data obtained from the video footage. In the previous section we pointed out the focus of such a comparison -- the evolution of seating entropy in time. 

In this section we present the models in the order of their creation. For each model we introduce a  keyword that  is used to address the model. The evolution trajectory of the real world data is marked as \texttt{real}. All models work in a greedy manner. An arriving group chooses a place  to sit and remains seated there.  The models are following:

{\small
\begin{itemize}
	\item \emph{Random seat selection} (\texttt{random}) -- This model tries to implement the case when participants actually do not care about choosing a position and rather select an arbitrary empty seat. Hence, a group selects randomly a compact bunch of empty seats in one row where it can fit. 
	\item \emph{Maximization of personal space} (\texttt{max}) -- A group simply chooses a seating that maximizes the minimum Manhattan distance to other seated participants. Even though it does not seem that majority of participants select a seat in this manner this model is implemented for the sake of contrast and comparison. Measuring the minimum distance for a group of participants is equivalent to measuring the distance for each group participant separately and then taking minimum distance. 
	\item \emph{Personal space selection} (\texttt{space}) -- This model loosely incorporates the findings from the experiment data.   A group tries to find an empty space where the minimum Manhattan distance to others is somewhere between 2 and 4 (including the bounds). If there is no such spot it tries to find a place at even larger distance. If such seat cannot be found, the group selects a random available space.  
	\item \emph{Simple space selection} (\texttt{simple}) -- The following model is a combination of the first model and a simplification of the previous model. In this setup a group does not care about seating too much. However, it rather chooses some space where it has at least some amount of personal space, i.e.,  basically one empty seat is left around the group. This is operationalized by the group having a minimum Manhattan distance to all participants greater than 2. If there is no such spot then a seating is chosen randomly. 
	\item \emph{Center of mass selection} (\texttt{center}) -- A group considers all available spaces that have minimum Manhattan distance to others greater or equal to 2. From such available positions it selects randomly among the ones that are closest to the center of the mass of already seated participants.  If there is no such space available, then the group chooses its spot randomly.   
\end{itemize} }

\begin{figure}
	\centering
	\includegraphics[width=7cm]{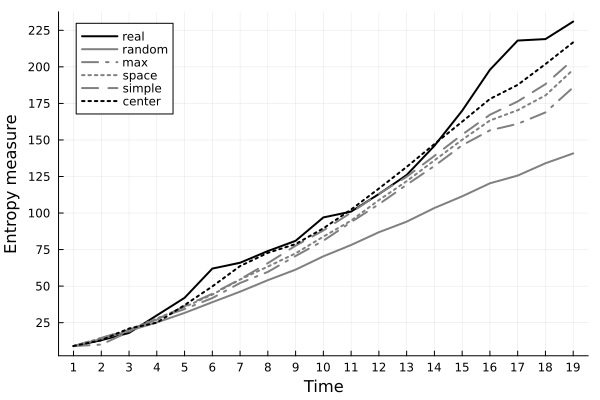}
	\caption{Evolution of entropy in time for each of the predefined models and the real data. For each model the trajectory is averaged over 1,000 independent runs of the model.} 
	\label{fig:comparison}
\end{figure}

In Figure \ref{fig:comparison} we show comparison of the models. The last method (\texttt{center}) incorporating the information about the center of mass  seems to resemble the actual real entropy evolution in the closest way. The model with simple minimization of distance to the center of mass seems  The model is the only one that  follows the "bump" of the real data in between the time slots 7 and 9.

Random selection model (\texttt{random}) and maximization model (\texttt{max}) provide the lowest resemblance to the real data. This supports the idea that participants neither do not select their seating positions randomly nor do they maximize distance from others . The simple variant with embedding the idea of personal space (\texttt{simple}) seems to be much more accurate.  The most complex method (\texttt{space}) is outperformed by the simpler method (\texttt{simple}). Such an observation might be consistent with Occam's razor principle. 

\section{Summary}   
 Several entities could disseminate in the physical space -- information, moods, memes, gossips. If we adhere to the previous studies of human territoriality and privacy we could understand personal space to be an entity that also disseminates in the physical space. Personal spaces of people intervene and force each other to form various patterns. Some of the patterns can be observed during the process of seating arrangements.  Hence, the main focus of the work was studying the process of emergence of seating configurations in rectangular auditorium. Especially, when participants arrive in smaller groups.

In this work we utilized a time lapse footage of a lecture hall being filled up and questionnaires with seating examples in order to be able to form models of emerging seating patterns. The model where seat selection is based not only on reasonable distance to others but also on distance to the center of the mass of already seated participants seems to be the most plausible. Its evolution of auditorium entropy in time resembles most closely the entropy of data obtained from the real footage. 

However, the data we used have limitations. The video footage is only one-time example of such process. Therefore, it does not capture the vast possibilities of seating arrangements. The questionnaire might suffer from being of artificial nature and not reflecting the true basis of seat selection process entirely. 

In the  real seat selection process many factors can play a role. We discussed them in the second section of this paper. Some of them could be neglected by limiting the modeled situations, some of them cannot be properly modeled (e.g., the attraction among participants), some of them need to be further employed in future models.      

In future work we intend to capture our own footage of emerging seating arrangements. The method of using questionnaires also has its advantages, hence we intend to obtain much larger data collection with more refined test configurations. Especially, it is worth testing the current hypothesis about the influence of the center of the mass of already seated participants. Also some other test configurations are needed to enlighten the behavior of participants when an auditorium is nearly full.  It would also be beneficial to enable a verbal feedback of participants for individual configurations. 

Another interesting research direction is exploration of various measures of distance and entropy. In this work we mostly used the Manhattan distance measure and our simple custom entropy measure. Exploration of other variants might bring even more realistic approach to how participants perceive  seating arrangements and also better model comparison.  

One particular research question attracts us. A real seating arrangement is formed locally by people gradually coming to an auditorium. The final seating arrangement is obtained by a process we could call "crowd social computing". We can  understand participants and their collectivity as a computational unit that is able by its own gradual self-organization to reach a certain configuration. Our question is how close (or how far) such a configuration is from a patterns computed globally in advance, i.e., by means of optimization. That is because several optimization areas deal with similar tasks, e.g., rectangle packing or, if we perceive the auditorium as a set of individual rows, bin packing. 

Last but not least, results of studying the seat selection process can be applied to design of lecture or concert halls. Especially, to enable better use of their space and to help participants become more comfortable and concentrated.

\subsubsection{Acknowledgements} 
The work of the first author was supported by Charles University, project GA UK No. 234322. The work of the second author was supported by the Czech Science Foundation project 23-07270S.

 \bibliographystyle{splncs04}
 \bibliography{biblio}
%Author, F.: Article title. Journal \textbf{2}(5), 99--110 (2016)
%Author, F., Author, S.: Title of a proceedings paper. In: Editor,
%F., Editor, S. (eds.) CONFERENCE 2016, LNCS, vol. 9999, pp. 1--13.
%Springer, Heidelberg (2016). \doi{10.10007/1234567890}
%Author, F., Author, S., Author, T.: Book title. 2nd edn. Publisher,
%Author, A.-B.: Contribution title. In: 9th International Proceedings
%on Proceedings, pp. 1--2. Publisher, Location (2010)
%LNCS Homepage, \url{http://www.springer.com/lncs}. Last accessed 4
%\end{thebibliography}
\end{document}